# On Physical Web models

Manfred Sneps-Sneppe
Ventspils International Radio Astronomy Centre
Ventspils University College
Ventspils, Latvia
manfreds.sneps@gmail.com

Dmitry Namiot
Faculty of Computational Mathematics and Cybernetics
Lomonosov Moscow State University
Moscow, Russia
dnamiot@gmail.com

*Abstract*— The Physical Web is a generic term describes interconnection of physical objects and web. The Physical Web lets present physical objects in a web. There are different ways to do that and we will discuss them in our paper. Usually, the web presentation for a physical object could be implemented with the help of mobile devices. The basic idea behind the Physical Web is to navigate and control physical objects in the world surrounding mobile devices with the help of web technologies. Of course, there are different ways to identify and enumerate physical objects. In this paper, we describe the existing models as well as related challenges. In our analysis, we will target objects enumeration and navigation as well as data retrieving and programming for the Physical Web.

*Keywords—network proximity; Physical Web; Bluetooth; Wi-Fi*

## I. INTRODUCTION

The Physical Web is a term that describes the process of presenting everyday objects on Internet [1]. It aims to offer users the way to perform their daily tasks at using surrounding objects, as soon as these objects are smart and remotely controllable. It is the main idea - perform everyday tasks depending on the surrounding physical objects. For the physical objects, we should pay attentions to the existence and the states. Of course, the states of objects (measurements) could have some history (e.g., time series of measured value). So, the key moment here is the introduction of some metric (metrics) for the physical objects. And of course, any introduced metrics should be measurable. We should suggest the easy (cheap) way to measure introduced attributes. On practice, any model for the Physical Web is just a set of metrics as well as use cases for their deployment. The use cases let us navigate and control physical objects in the world surrounding mobile devices.

The first question for any metric is the way to enumerate physical objects. For example, we can mention well-known QR-codes [2] as a typical example of enumerating. Another widely used approach here is the deployment of RFID technology [3], or, more recently, wireless tags [4].

 Wireless tags can support standard protocols like Bluetooth (Bluetooth Low Energy) and Wi-Fi. So, for mobile devices (mobile users) the detection of tags is actually the detection of wireless nodes. It solves the above-mentioned problem with the measurability. And there are two important moments. This detection could be performed programmatically. The modern mobile operational systems (iOS, Android) provide application program interfaces (APIs) for such tasks. Secondly, in this approach other mobile devices can play a role of the tag too. For example, a mobile phone could be presented as a Wi-Fi access point or Bluetooth node in the so-called discoverable mode (Figure 1).

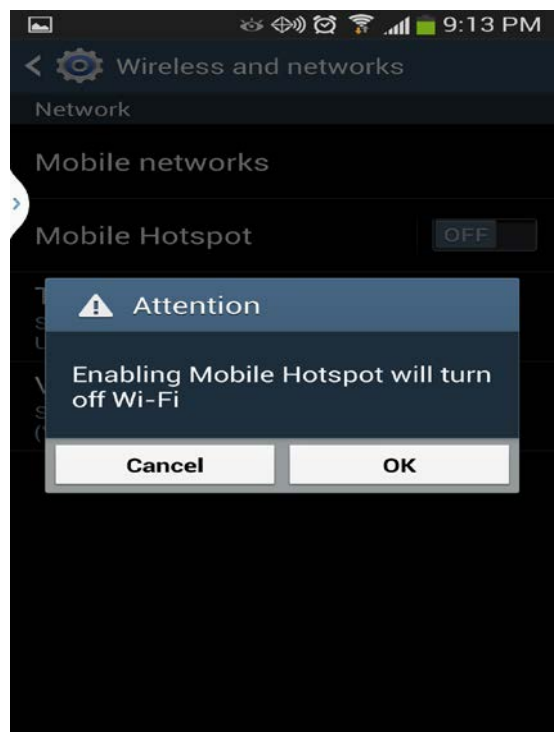

Fig. 1. Wi-Fi access point on the mobile

The option opens the way for the very interesting line of use cases. If we link (associate) some data with the visibility of such node, data availability will follow to the mobile device. And any movement for the device will cause the movement for data availability too. Think, for example, about some classified data, published by the owner of mobile hot-spot. In this case, the visibility for his announce depends on the current location of the mobile phone. In other words, his announce will be visible for the readers who are currently nearby the author. It can increase the conversion rate in the commercial applications, for example.

In general, this model is so-called network proximity [5]. The network proximity here describes data models based on the detection of surrounding network nodes.

In this paper, we would like to discuss several approaches for building mobile computing systems based on the detection of physical objects via network proximity. The classical models of interaction with physical objects are a subject of Internet of Things (Web of Things) [6]. In our paper, we will mostly discuss the services which could be associated with the presence of surrounding physical object. The fact that ant particular object is "visible" for the mobile user can trigger some actions and/or change the output for mobile services. It is so-called ambient intelligence (AMI) [7]. AMI is a paradigm which it aims multidisciplinary development physical environments where different electronic objects intelligently respond to the presence of people [8]. AMI targets the creation of sensitive, adaptive electronic environments that respond to the actions of persons and objects and cater for their needs. AMI approach includes the entire environment (all physical objects) and links (associates) it with human interaction. As a result, we can expect an extended and more intuitive interaction, enhanced efficiency, increased creativity, etc.

Note, that such services do not always include any form of two-way (or even one-way) data exchange with the physical objects. In the most cases, it is enough to detect and identify the object. In other words, we can deal with the proximity information only.

The proximity is a very conventional way for context-aware programming in the mobile world. There are many practical use cases, where the concept of the location can be replaced by that of proximity. Proximity can be used as the main formation for context-aware browsers [9]. The idea is to allow automatic download of Web pages, and even automatic execution of web applications, on user's own mobile device. The web resources are not simply pushed on the mobile device; rather, they are selected on the basis of the context the user is in: context data are used to build a query sent to an external search engine, which selects the most relevant web content [10]. In our early projects, the context-aware browser was used for dynamically generated output in mobile information services [11].

The usage of network proximity for context-aware systems is very transparent. At his moment, network modules are most widely used "sensors" for mobile phones. All modern smartphones have Wi-Fi (Bluetooth) modules. So, Wi-Fi (Bluetooth) related measurements are included in standard interfaces of mobile operating systems. At least, the above-mentioned measurements include the visibility for network nodes (presence or existence of tags). As additional attributes, we can use signal strength and the special information provided, for example, by Bluetooth low energy tags. By the definition, the distribution of Wi-Fi and Bluetooth signals is limited. So, if any Bluetooth node is visible from a mobile device (a mobile phone, for example), then this device is somewhere nearby that node (it is so-called Bluetooth distance). The same is true for Wi-Fi access point. And this proximity information (network proximity) is the typical context information which can replace location data in many services. There are two main reasons for this replacement. At the first hand, we can target here all indoor application [12]. Obtaining GPS (Global Positioning System) data indoor is not reliable and sometimes even impossible. In the same time, most of the offices usually have plenty of wireless nodes. The second reason for this replacement is the above-mentioned mobile wireless node (e.g., Wi-Fi access point right on the mobile phone), when our context information will follow to the moved object.

As we have mentioned in our previous papers, for network proximity-based context-aware applications, any existing or even especially created Bluetooth node could be used as a presence sensor that can play the role of a trigger. This trigger can open access to some content, discover existing content, as well as cluster nearby mobile users [13, 14].

The rest of the paper is organized as follows. In Section II, we discuss QR-code based systems. In Section III, we discuss iBeacons (Apple) and Eddystone (Google Physical Web). In Section IV, we describe our Bluetooth Data Points model.

## II. QR-CODE FOR THE PHYSICAL WEB

QR-codes are the well-known approach for attaching some data to physical objects (Figure 2).

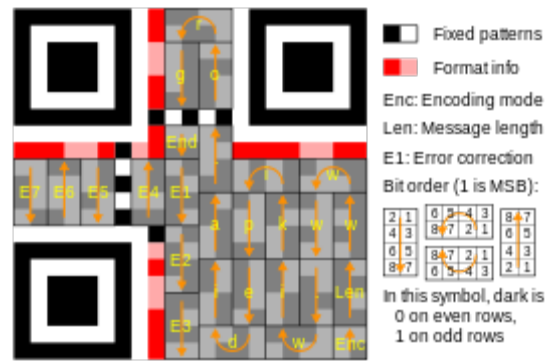

Fig. 2. QR-code format

QR-code (Quick Response Code) is the trademark for a type of matrix barcode (or two-dimensional barcode). It is a machine-readable optical label that contains information about the item to which it is attached [15]. Typically, a mobile phone (mobile user) is used as a QR code scanner. It converts the code to some useful form (URL, phone number, plain text). In general, 2D barcodes encode some text. But in many cases, that text can represent some specified things. In the most often usage, 2D barcodes encode text that represents some URL, like "http://some_domain.com/". Technically, it is a special use case for text representation, where URL's pattern is recognized by the software. The recognized URL obviates the need for a user to type it into a web browser. QR-code reader can recognize an URL and open it in the browser.

In this connection, we can mention, for example, context-aware QR-code reader [16]. It is a typical mashup. This context-aware QR-code scanner is based on the modified version of open sourced scanner Zxing from Zebra Crossing. Context-aware QR-code reader keeps the basic processes of scanning and recognizing as they are, but just adds some parameters on the final stage. In other words, the customized QR-code scanner will replace encoded value

http://some_domain.com/

with

http://some_domain.com?list_of_parameters

And this list of automatically added parameters will describe our context. For example, let us see the QR-code deployment for some indoor retail application, provided mobile coupons. QR-codes will let mobile visitors download an appropriate coupon. URLs are the perfect instrument for this task. Technically, we can prepare a separate QR-code for the each existing department (for the each existing discount program). But this process could be costly and difficult to maintain. Context-aware QR-code reader can automatically add context information to any encoded URL. This context information could be so-called network fingerprint. Shortly, it is a list of the "visible" wireless nodes. This idea lets us use the same generic URL (and generic QR-code too) across all our installations. The URL in QR-code points to some CGI-script which can proceed HTTP GET parameters in the request and respond accordingly.

Note, that the QR-code here is just a visible (and automatically recognizable) presentation. Of course, we use different forms. For example, Reality Editor from MIT [17] uses another form (Figure 3)

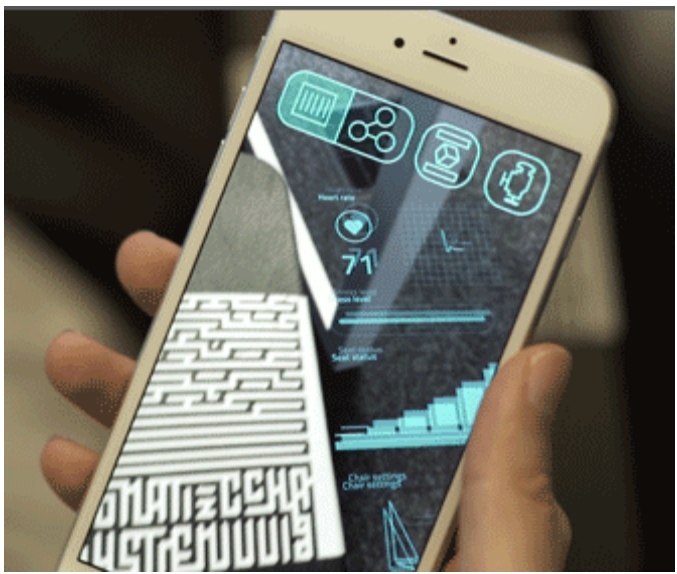

Fig. 3. MIT Reality Editor

The Reality Editor is a new kind of tool for empowering you to connect and manipulate the functionality of physical objects. And objects are recognized by the visible labels.

### III. IBEACONS AND EDDYSTONE

The iBeacon is a wireless tag (beacon), based on Bluetooth Low Energy (BLE) standard [18]. Shortly, any beacon is set to transmit a set of numbers several times per minute, so that any mobile device with BLE support nearby can detect it. The beacon's repetitive transmission is called also as "advertising". The BLE standard specifies a structure for the data that must be transmitted. An application on a mobile phone can then detect this parcel of information, unpack it, and use it for providing context-aware services.

The above-mentioned advertising includes a unique ID for a tag and two application-dependent numbers (so-called minor and major) [19].

As per Apple's manual, a proximity universally unique identifier (UUID) is 16 Bytes, and major and minor codes are 2 bytes each. The common usage for UUID is the identification of a place. For example, it could be a particular shop, café, etc. Major and minor codes could be used to a description of an area within a physical space associated with the above-mentioned UUID. For example, a retailer might use the major and minor code to identify, respectively, a given retail store and a specific shelf, where a beacon will be placed.

On iOS, a given application can scan for up to 20 tags (proximity UUIDs). It is, probably, one of the biggest limitations for iBeacons technology. The mobile application should statically declare UUIDs for the tags in questions [20]. For a mobile application, this declaration lets register be notified if a Beacon with a given UUID comes within range (or goes out of range) of the device [21]. From the notification, a mobile application can obtain minor and major codes and they can then be used to uniquely identify a given beacon.

The application can then use this data to decide what action to take, if any, when the beacon is detected. This detection could be included into the application itself. But more often (if not almost always) in should be done in tandem with a cloud service. So, the cloud service for data processing is a part of this story.

Beacons could be placed anywhere where potential users might wish to either trigger some form of action in a mobile application, or have that application log the fact that it came near to the beacon. For example, CES expo scavenger hunt was tested at CES recently. The idea of the scavenger hunt was to encourage attendees to explore the event. People downloaded the CES mobile app onto their iOS or Android devices and then looked to find iBeacon badges throughout the conference venue and win a prize. [22].

The main technological problem, by our opinion, is the need for the static description of observer tags. Of course, the underlying system (iOS) can read data from all tags in the proximity, but dispatches only some of them to an application. It means that the only one company (Apple) has the whole picture.

Google comes with the own protocol for BLE [23]. Eddystone is the protocol specification that defines a Bluetooth low energy (BLE) message format for proximity beacon messages. It describes several different frame types that may be used individually or in combinations to create beacons that can be used for a variety of applications. At this moment, we can see the following frames (types of data) in the protocol:

Eddystone-UID: an opaque, unique 16-byte Beacon ID composed of a 10-byte namespace ID and a 6-byte instance ID. The Beacon ID may be useful in mapping a device to a record in external storage. The namespace ID may be used to group a particular set of beacons, while the instance ID identifies individual devices in the group. It is an analog for a minor/major pair in iBeacon from Apple.

Developers typically can use the namespace ID to signify own company or organization, so they know the owner for a beacon.

Eddystone-URL: it is a URL in a compressed encoding format. Once decoded, the URL can be used by any client with access to the Internet. It is a link to the Google Physical web, we will discuss below. Semantically, it is an analogue for the above-mentioned URL, encoded with QR-code.

Eddystone-TLM it is a frame which broadcasts telemetry information about the beacon itself. This information includes battery voltage, device temperature, and counts of broadcast packets. It contains the packet version (always a one-byte value of 0 for now), the beacon temperature (2 bytes), the beacon battery level (2 bytes), the number of seconds the beacon has been powered (2 bytes) and the number of "PDU" packet transmissions the beacon has sent (2 bytes.)

Actually, the general idea (pattern) is the same as with the "classical" iBeacons. Tags broadcast some ID, an application uses ID for getting data from the cloud. URL here is just a special case of ID. We can simulate URL transmission just by mapping tag's ID to some URL in the cloud-based data store. Anyway, with obtained URL application should get access to the Internet for obtaining data.

Google provide Proximity Beacon API for setting attachment (data associated with) for BLE tags [24]. This API lets developers register tags, associate data with tags (add attachments in Google's terms), retrieve data from tags (retrieve attachments) and monitor beacons. Technically, it is the same as for iBeacons, except telemetry data.

The tag itself has got a rich set (comparing with iBeacons) of attributes: advertised ID (required), current status, expected stability, geo-coordinates (latitude, longitude pair), ID for Google Places [25], indoor floor level and text description.

The attachment (data) is a string up to 1024 bytes long. It could be a plain string, JSON data or even encoded binary data. Attachments are stored in Google's scalable cloud. So, dislike iBeacons, Eddystone originally has got back-end support.

There is also a very important remark for the development: on Android platform, it is possible to obtain information about all "visible" tags.

Eddystone is a part of *Nearby* API [26]. *Nearby* API lets create features based on proximity. As per Google's documents, *Nearby* API exposes simple publish and subscribe methods that rely on proximity. Your application publishes a payload that can be received by nearby subscribers. On top of this foundation, developers can build a variety of user experiences to share messages and create real-time connections between nearby devices.

There are two use cases: Nearby Messages API and the Nearby Connections API. Nearby uses a combination of Bluetooth, Wi-Fi, and inaudible sound (using the device's speaker and microphone) to establish proximity (Figure 4).

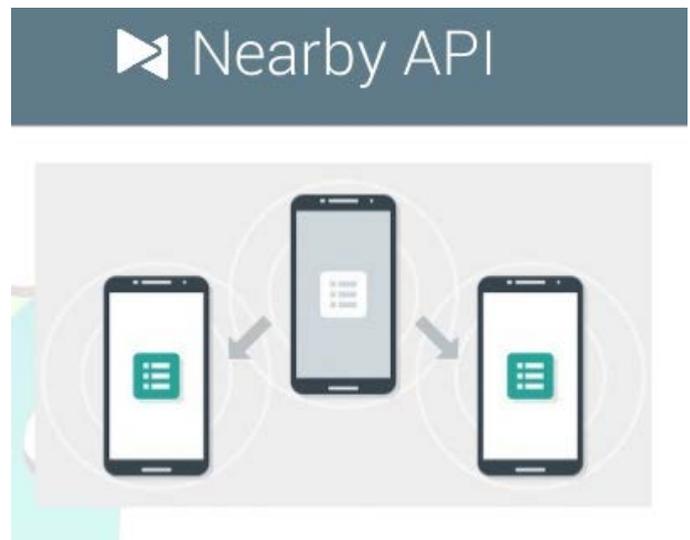

Fig. 4. Nearby API

Google Physical Web project is an example of integration Web technologies and physical world. Actually, it is part of a more generic problem: how to integrate Internet of Things and web technologies [27]. As per Google's vision, the Physical Web is an example of discovery service. In this model, a smart Physical Object broadcasts relevant URLs that any nearby device can receive. This simple capability can unlock exciting new ways to interact with the Web [28].

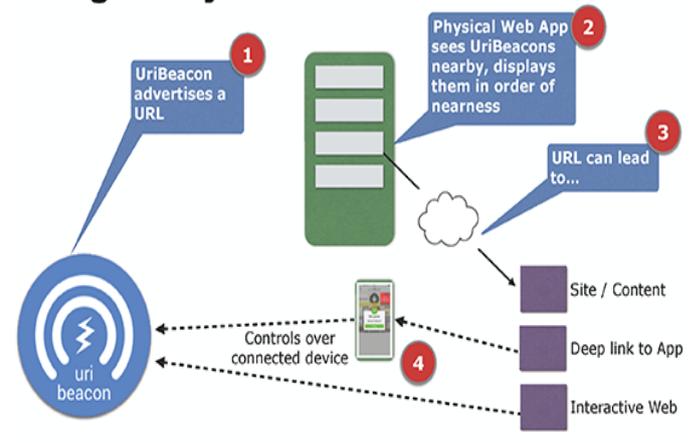

Fig. 5 The physical web

Figure 5 illustrates the basic idea. Actually, the physical objects here are (in the most cases) the same Bluetooth tags. In the current implementation, URL broadcast method involves a Bluetooth broadcast from each tag. The user's phone can obtain this URL without connecting to the beacon. As per Google, this ensures the user is invisible to all beacons, meaning a user can't be tracked simply by walking past a broadcasting beacon. For example, in passive Wi-Fi monitoring, silent users could be tracked automatically. As per Google, this was very much by design to keep user's silent passage non-trackable. But it is assumed also, that URL detection should be performed automatically, in the background. Of course, once the user opens a URL (does click on a URL), he is known to that website. With this solution, Google mostly follows to iBeacon usage (deployment model). Application on the mobile device automatically discovers nearby objects, obtains associated data (URLs in this case) and pushes this information to the user.

It is the true push approach, because both iBeacons and EddyStone are using push notifications, supported by mobile OS [29]. Notification service is a popular functionality provided by almost all modern mobile OS (iOS, Android, etc). To facilitate customization for developers, mobile platforms support highly customizable notifications. But there are two main issues with push notifications. Firstly, it is disruptive for user interfaces. Even the loyal subscriber could be not intended to receive messages at this moment. Secondly, the third-party push notification customization may allow an installed trojan application to launch phishing attacks or anonymously post spam notifications [30]. So, why do not switch to browsing mode instead of push notifications? Mobile applications may still obtain tag's data automatically, prepare some context-aware information, but show it only when a user directly requests it. It is exactly how a browser works. We must see the direct intention from mobile users to obtain nearby data. In this case, our application should form dynamically a web page (like CGI-script in the web) and show it to the user.

We can mention here a very simple approach for creating the Physical Web for any Bluetooth/Wi-Fi device. What if we define a SSID (name) for Wi-Fi access point (Bluetooth node) as some URL? SSID for wireless node is broadcasted. This broadcast is similar (semantically) to the advertising in terms of the Physical Web. Of course, this setup could be done programmatically. And mobile application (programmatically also) can get a list of available Wi-Fi access points (Bluetooth nodes in the so-called discoverable node). This list includes SSIDs (URLs). So, it is a typical Physical Web, even without the advertised tags like iBeacons or EddyStone. Wi-Fi access point (Bluetooth node in the discoverable mode) could be set programmatically (it is true for Android) right on the mobile phone. So, any smart phone could be turned into a Physical Web tag and provide advertising in the form of some URL.

In the more generic form, this approach could be described as Beacon stuffing [31]. It is a low bandwidth communication protocol for IEEE 802.11 networks that enables Wi-Fi access points to communicate with clients without association. This enables clients to receive information from nearby access points even when they are disconnected, or when connected to another access point. Originally, this scheme was developed for Wi-Fi as complementary to the 802.11 protocol. It works by overloading 802.11 management frames while not breaking the standard. By the similar manner, this scheme will work for Bluetooth [32]. Actually, our own idea of Bluetooth Data Points has been inspired by Beacon Stuffing.

## IV. BLUETOOTH DATA POINTS

Bluetooth Data Points (BDP) [33] let us turn any Bluetooth node into a tag. The main idea behind BDP is to link (associate) user-defined data with existing wireless networks nodes. In this process, we can use any existing nodes as well as especially created elements. Originally, the project targets Bluetooth nodes in the discoverable mode, but the same principles will work for Wi-Fi access points too. The associated data for a particular wireless node is a direct analogue of the above-mentioned tag's attachment. The main difference is the definition (the description) for a tag. BDP is based on the idea of "zero scene preparation". We do not need to place and configure tags. For example, any mobile users should be able to create (open) Bluetooth node right on the own mobile phone, associate some data with this node and so, make them available for other mobile users in the proximity. Figure 6 illustrates this idea. As an existing node, we see here Bluetooth node in the car. Many modern cars nowadays are actually Bluetooth nodes. Car's owner can attach data to the own node. Other mobile users in the proximity can "see" Bluetooth node and use its identification (SSID, MAC-address) as key for obtaining associated data from the cloud. So, rather than directly advertise some encoded URL, BDP's "tags" advertise own SSIDs (as usually). And any SSID is a key for obtaining data from the cloud.

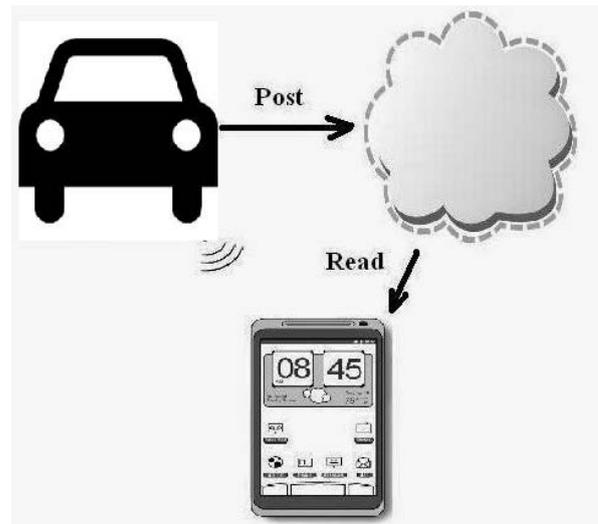

Fig. 6. BDP data flow.

It is so-called hyper-local data concept. Data not only associated with some local information but could be prepared locally also. Instead of the car (Bluetooth node in the car) in Figure 6, we can use just another mobile phone. A Bluetooth node (a tag in BDP conception) could be created programmatically. The publisher can attach some data to it via BDP application too. So, from the publisher's point of view, one mobile application is enough for creating a new data channel. And the same mobile application (in browsing mode) could be used for reading data in the proximity.

The simplest use case is the above-mentioned mobile classified. A mobile user creates an advertising (announce), links it to the wireless node on the own mobile phone and so, makes it available for reading for other mobile users in the proximity. If the mobile phone (the mobile tag) is moved, all associated data will be "moved" too. Data are not associated with latitude/longitude pair (as in geo-location systems), but with ID of the wireless node. Data are visible in the proximity of the node (in the proximity of the author for classified) only.

Google does not describe the above-mentioned database for beacons data attachments. BDP uses the classical key-value model for data. It is based on Open Source Apache Accumulo distributed key-value store and custom solution for the cache.

## V. CONCLUSION

In this paper, we discuss several models for the physical web. As the most promising approach in this area, we propose to use network proximity. As the main result, we present our list of requirements to the flexible solution, based on the wireless tags.

By our opinion, the Physical Web approach should support software-based tags. As per network proximity, it should be possible to define tags and linked data with existing wireless infrastructure as well as directly with mobile devices. iBeacons support software-based tags, Google's Eddystone does not support them. Our BDP approach is completely based on software-defined tags. In our review, we do not reject the idea of using dedicated hardware tags. But we think that software-based systems are much more flexible, cheaper and allow much more use cases.

## ACKNOWLEDGMENT

We would like to thank prof. Vladimir Vishnevsky for the valuable discussions.